\documentclass[12p]{iopart}
\usepackage[utf8]{inputenc}
\usepackage{amsfonts}
\usepackage{amssymb}
\usepackage{graphicx}
\usepackage{url}
\usepackage{xcolor}

\newcommand{\red}[1]{{#1}}

\begin{document}
\title{Exact sum rules with approximate ground states}

\author{Calvin W. Johnson}
\ead{cjohnson@sdsu.edu}
\address{
 San Diego State University, San Diego, California, USA
}
\author{Ken A. Luu}
\address{
 San Diego State University, San Diego, California, USA
}

\author{Yi Lu }
\ead{luyi@qfnu.edu.cn}
\address{Qufu Normal University, Shandong, China}

\date{\today}

\begin{abstract}
Electromagnetic and weak transitions tell us a great deal about the structure of atomic nuclei.  Yet modeling transitions can be difficult: it is often easier to compute the ground state, if only as an approximation, than excited states. One alternative is 
through transition sum rules, in particular the non-energy-weighted and energy-weighted sum rules, which can be computed as  expectation values 
of operators.  We investigate by computing  sum rules for a variety of nuclei, comparing the numerically exact full configuration-interaction
shell model, as a reference, to Hartree-Fock, projected Hartree-Fock, and the nucleon pair approximation. 
These approximations yield reasonable agreement, which we  explain by prior work on the systematics of transition moments.
\end{abstract}

\maketitle



\section{Introduction}

Atomic nuclei are complex many-body systems displaying a wide variety of phenomena, such as rotational and vibrational bands and superfluid behavior.  In order to go beyond  
energy spectra to probe wave functions, one must look at other observables. Static properties such as radii and multipole moments
give  insight, but to extract dynamical information one turns to transitions.  Electromagnetic and
weak transitions are particularly useful, as such transition can be appropriately computed in perturbation theory \cite{BG77}.  

Transitions require the explicit or implicit calculation of excited states. 
Some powerful many-body methods, however, such as density-functional theory\cite{baldo2008kohn,stoitsov2010nuclear,erler2013energy}, 
coupled cluster methods \cite{hagen2010ab,hagen2014coupled}, and Monte Carlo 
methods \cite{PhysRevC.48.1518,koonin1997shell,pieper2005quantum,pieper2001quantum}, are best at computing the ground state, as a natural consequence of the variational theorem. 
Extensions to excited states are often much more
computationally expensive than for the ground state. 

One can get around this limitation by using sum rules \cite{bohigas1979sum,lipparini1989sum,ring2004nuclear,PhysRevC.89.064317,PhysRevC.97.034330}.  
  Let $\hat{\cal O}$ be some transition operator, and 
let $ \{ | n \rangle \}$ be eigenstates of a Hamiltonian $\hat{H}$ with energies $E_n$. The transition strength function is
\begin{equation}
S(E_i,E_x) = \sum_f \delta(E_x - E_f + E_i) \left | \langle f \left | \hat{\cal O } \right | i \rangle \right |^2,
\end{equation}
while sum rules are weighted integrals, or moments, of the strength function. The non-energy-weighted sum rule (NEWSR) is just
\begin{equation}
S_0(E_i) = \int dE_x \, S(E_i, E_x) = \sum_f \left | \langle f \left | \hat{\cal O } \right | i \rangle \right |^2,
\end{equation}
while the energy-weighted sum rule (EWSR) is
\begin{equation}
S_1(E_i) = \int dE_x \, E_x S(E_i, E_x) = \sum_f ( E_f - E_i) \left | \langle f \left | \hat{\cal O } \right | i \rangle \right |^2.
\end{equation}
Other moments are possible and are used, for example in calculating polarizabilities \cite{ring2004nuclear,PhysRevC.94.034317}, but these two are particular easy to compute, because one can rewrite them 
using completeness relations as expectation values:
\begin{eqnarray}
S_0(E_i) &=& \langle i | \hat{\cal O}^\dagger \hat{\cal O} | i \rangle, \\
S_1(E_i) &=& \frac{1}{2} \left  \langle i \left |   
\left [ \left [ \hat{\cal O}^\dagger, \hat{H} \right ], \hat{\cal O} \right ]
 \right  | i \right \rangle 
\end{eqnarray}
These expressions  depend only upon the completeness of the model space and, in the case of the EWSR, the behavior under Hermitian conjugation of the transition operator. 
Recently two of us derived and wrote a computer program for the matrix elements $\hat{\cal O}^\dagger \hat{\cal O} $ and 
$ \frac{1}{2} \left [ \left [ \hat{\cal O}^\dagger, \hat{H} \right ], \hat{\cal O} \right ]$ for general one-body transition operators, including nonscalar operators with 
nonzero angular momentum rank, and general 
$1+2$-body Hamiltonians  \cite{PhysRevC.97.034330}.  This allows one to efficiently compute the NEWSR and EWSR for many nuclei for arbitrary 
one-body transitions.  
{(Many transitions have, in principle, important corrections due to two-body currents \cite{PhysRevC.78.064002,PhysRevC.78.065501,PhysRevLett.107.062501}, but for calculations of medium and heavy nuclei with empirical interactions, 
effective one-body operators are still primarily used \cite{brown1987empirically,PhysRevC.53.R2602}. )}

While individual transitions are important for specific physical applications, such as astrophyical rates, collections of 
transitions carry much  more information. For example, relative E2 transitions tell us about the collectivity and deformation of a 
nucleus, and even play a role in identifying underlying algebraic structures.  These insights carry over to sum rules. 
For example, although we do not explore it in this paper, the Thomas-Reiche-Kuhn (TRK) energy-weighted sum rule for electric dipole 
transitions\cite{PhysRevA.60.262} gives an analytic results for local interactions--and thus deviations from TRK measure the momentum-dependence 
of, say, nuclear interactions \cite{towner1977shell,price1985electric}, while Gamow-Teller sum rules are sensitive to the smearing of the Fermi surface \cite{PhysRevLett.65.1325}, and so on. 
{In particular, these sum rules could be useful checks on calculations of electromagetic or weak transition strength functions when carried out in methods using 
only a small fraction of the model space, such as the Monte Carlo shell model \cite{honma2005effective}, the nucleon pair approximation \cite{PhysRevC.62.014315}, 
or the projected shell model \cite{tan2020novel}.  Because we demonstrate here that sum rules are generally robust even on approximate ground states, one can have confidence 
in such a cross-check.
}

Given the usefulness of transitions generally and sum rules in particular, and the relative ease of generating ground states by a 
variety of methods, in this paper we compare the sum rules from various many-body approximations to the ground state, against 
an `exact' ground state, here defined as a full configuration-interaction calculation.  We find generally good agreement, which we {analyze 
in Sec.~\ref{analysis}} by appealing to prior work on systematics of sum rules.  
{Our work is motivated by  the call of the authors of \cite{lipparini1989sum} to investigate sum rule methods beyond the mean-field; not 
only do we apply a general method, we present evidence} one can use  approximate ground states as a good test of
nuclear dynamics.

\section{Methods}

We work in a shell-model basis, that is, many-body states built from single-particle states with good angular momentum $j$ and parity $\pi$.  One must restrict the 
single-particle states, which leads to a finite many-body basis which can nonetheless be very large.  
We use two valence spaces: the $1s_{1/2}$-$0d_{3/2}$-$0d_{5/2}$ space, of $sd$-shell, with a frozen $^{16}$O core, 
and the 
$1p_{1/2}$-$1p_{3/2}$-$0f_{5/2}$-$0f_{7/2}$, or $pf$-shell, with a frozen $^{40}$Ca core. In the $sd$ shell we use 
the universal $sd$-shell interaction version B, or USDB \cite{PhysRevC.74.034315}, and in the $pf$ shell we use a modified $G$-matrix interaction version 
$1A$, or GX1A \cite{PhysRevC.65.061301,honma2005shell}. These are both `gold-standard' interactions which reproduce many data well. 
The Hamiltonian has one- and two-body parts, which, using fermion creation and annihilation operators $\hat{a}^\dagger, \hat{a}$, respectively, becomes
\begin{equation}
\hat{H} = \sum_{ij} T_{ij} \hat{a}^\dagger_i \hat{a}_j + \frac{1}{4} V_{ijkl} \sum_{ijkl} \hat{a}^\dagger_i \hat{a}^\dagger_j \hat{a}_k \hat{a}_l.
\label{ham}
\end{equation}
The operators are coupled up to an angular momentum scalar \cite{BG77}.  Aside from those restrictions, our codes can use any interaction, which is 
provided externally as a file. 

With these interactions in these model spaces we solve for the ground state wave function and its sum rules. The latter are, as discussed above, computed as expectation values of a
scalar operator which has the same form as the Hamiltonian (\ref{ham}).

\subsection{Full configuration-interaction}

Our benchmark is full configuration-interaction (FCI), sometimes called the interacting shell model \cite{BG77,towner1977shell,lawson1980theory,br88,ca05}, where one expands the wave function in a many-body basis $\{ | \alpha \rangle \}$:
\begin{equation}
| \Psi \rangle  = \sum_\alpha c_\alpha | \alpha \rangle.
\end{equation}
For our basis we use antisymmetrized products of single-particle states, or Slater determinants; or, technically, the occupation-representation
of Slater determinants: If $\hat{a}^\dagger_i$ is the creation operator for the $i$th single-particle state, then the occupation representation of 
an $A$-body Slater determinant is
\begin{equation}
\hat{a}^\dagger_1 \hat{a}^\dagger_2 \hat{a}^\dagger_3 \ldots \hat{a}^\dagger_A | 0 \rangle,
\label{SD}
\end{equation}
where $| 0 \rangle$ is the fermionic vacuum, or, equivalently, a frozen core. 

Because both total angular momentum $\hat{J}^2$ and the $z$-component $\hat{J}_z$ commute with 
our Hamiltonians, we choose many-body basis states with fixed eigenvalues of the latter, labeled as $M$. This is known as an \textit{M-scheme basis}, and is easily accomplished when using single-particle states $i$ with good angular momentum $j_i$ and $z$-component
$m_i$, so that the total value of $J_z$ of (\ref{SD}) is $m_1 + m_2 + m_3 + \ldots m_A$.   Other than fixing $J_z$, for FCI 
we take all possible Slater determinants.  In this framework it is easy to construct an orthonormal many-body basis, $\langle \alpha | \beta \rangle = \delta_{\alpha \beta}$. 
The code we use \cite{BIGSTICK,johnson2018bigstick} efficiently computes matrix elements of the Hamiltonian in this basis, $\langle \alpha | \hat{H} | \beta \rangle$, 
and then uses the Lanczos algorithm to find low-lying eigenstates \cite{ca05,Lanczos}.  A Lanczos-like algorithm can also be applied to find the EWSR \cite{PhysRevC.89.064317}.

While FCI can find the numerically exact ground state, it is limited by the basis dimension.  Configuration interaction calculations are possible up to dimensions of  20 billion $M$-scheme 
states;  the largest dimensionalities requires a supercomputer and highly parallelized code.   It is this limitation that leads us to seek out other methods, whether exact (such as coupled-clusters) 
or approximate for computing sum rules.

All of the FCI calculations described here were carried out on a 32-core workshop with the largest cases taking a few hours.

To compare against our FCI results, which, we again remind the reader, are numerically `exact' given a fixed space, we consider three approximations: Hartree-Fock, angular-momentum projected Hartree-Fock, and the 
nucleon pair approximation.

\subsection{Hartree-Fock and projected Hartree-Fock}

Fortunately, we can carry out both Hartree-Fock  (HF) and angular-momentum projected Hartree-Fock  (PHF) in a shell-model framework, that is, using the exact same single-particle 
space and interaction matrix elements as used in our FCI calculations. Our Hartree-Fock code minimizes 
$\langle \hat{H} \rangle $ for an arbitrary Slater determinant $| \Psi \rangle$.  In particular, we redefine the single-particle basis 
by an $N_s \times N_s$ unitary transformation, where $N_s$ is the number of single-particle states,
\begin{equation}
\hat{c}^\dagger_a = \sum_i U_{ia} \hat{a}^\dagger_i,
\label{Utransform}
\end{equation}
 (actually an orthogonal transformation, as the only restriction we  impose is to force $U_{ia}$ to be real), and then let 
 \begin{equation} 
 | \Psi \rangle = \hat{c}^\dagger_1 \hat{c}^\dagger_2 \ldots \hat{c}^\dagger_A | 0 \rangle.
 \end{equation}
In this case, it is easy to represent the Slater determinant as a rectangular matrix ${\Psi}$, which for $N_p$ particle is given by $N_p$ columns, each of length $N_s$,  of the matrix $\mathbf{U}$ in 
Eq.~(\ref{Utransform}).  We have separate proton and neutron Slater determinants.  The formalism is straightforward  \cite{PhysRevC.48.1518}, if not widely disseminated, and is the basis for the path-integral formalism for the nuclear shell model \cite{PhysRevC.48.1518,koonin1997shell}
as well as the subsequent Monte Carlo shell model \cite{PhysRevLett.77.3315}.
One can compute $\langle \hat{H} \rangle$ for any ${\Psi}$ and then vary the elements of $\mathbf{U}$  to minimize  \cite{SHERPA}.   It is important to note that 
we find the HF minimum by using gradient descent \cite{ring2004nuclear} rather than diagonalizing the effective one-body Hartree-Fock Hamiltonian.  Gradient descent leads to much 
improved results, especially for odd-$A$ and odd-odd nuclides.

We can take any Slater determinant, including HF states, and project out states of good angular momentum. An arbitrary state $| \Psi \rangle$ can be expanded 
as sum of states with good angular momentum quantum number $J, K$, where $J$ is the total angular momentum and $K$ the $z$-component in the intrinsic frame:
\begin{equation}
| \Psi \rangle = \sum_{J,K} c_{J,K} | JK \rangle.
\end{equation}
Let $\hat{P}^J_{MK} $ be a projection operator that projects out a state of good angular momentum $J$ and $z$-component $K$ but rotated to $z$ component $M$.
In the standard approach one accomplishes this by an integral  \cite{ring2004nuclear}, but we perform the projection by solving a set of linear algebra equations \cite{PHF1, PHF2}.  Then one 
gets the angular-momentum projected Hamiltonian and overlap kernels, respectively:
\begin{eqnarray}
H^J_{MK} \equiv \langle \Psi | \hat{H} \hat{P}^J_{MK} | \Psi \rangle, \\
N^J_{MK} \equiv \langle \Psi | \hat{P}^J_{MK} | \Psi \rangle, 
\end{eqnarray}
and solve the generalized eigenvalue problem,
\begin{equation}
\sum_K H^J_{MK} g^J_K = E \sum_K N^J_{MK} g^J_K.
\end{equation}
We do this by instead constructing  $\tilde{H}^J = (N^J)^{-1/2} H^J (N^J)^{-1/2}$ (we carry out the inversion by spectral decomposition of the norm kernel, that is, diagonalizing it, allowing us to separate out 
singular or near-singular eigenvalues) and diagonalize  $\tilde{H}^J$ to get eigenvectors $\tilde{g}_K^J$.  These matrices are all of small dimension, of order 10 or so.

Evaluating the expectation value of a scalar $1+2$-body operator $\hat{\Omega}$ is then straightforward. We compute
\begin{equation}
\Omega^J_{MK}  \equiv \langle \Psi | \hat{\Omega} \hat{P}^J_{MK} | \Psi \rangle
\end{equation}
then transform $\tilde{\Omega}^J = (N^J)^{-1/2} \Omega^J (N^J)^{-1/2}$, and finally, using the eigenvectors of $\tilde{H}^J$, compute 
\begin{equation}
\langle \hat{\Omega} \rangle = 
\sum_{K K^\prime} \tilde{g}^{J*}_K \tilde{\Omega}^J_{KK^\prime} \tilde{g}^J_{K^\prime}.
\end{equation}
This can be done for any state that comes out of the generalized eigenvalue problem, but here we only apply it to the ground state.

In these model spaces, the HF calculations takes a few seconds and the PHF under a minute on a modest laptop.

\subsection{Nucleon pair approximation}
The nucleon pair approximation (NPA) starts from collective nucleon pairs in  occupation space
\begin{equation}
\hat{A}^\dagger_{r,m} = \sum_{ab} y(ab,r) [\hat{a}_a^\dagger \otimes \hat{a}_b^\dagger ]_{r, m},
\label{NPA-pair}
\end{equation}
where $a,b$ label single particle orbits with good angular momentum, ``$\otimes$" means tensor coupling so that the nucleon pair has good angular momentum $r$ and $z$-component $m$, and $y(ab,r)$ are the structure coefficients to allow for collectivity in the pair.

Usually only S, D, G, I pairs (i.e. $J=0,2,4,6$) are used, because in these low-angular-momentum pairs, two nucleons' wavefunctions overlap the most.
As the nucleon-nucleon force is mostly short-range and attractive, nucleons are energetically favored by large wave function overlap, which in turn occurs for low 
total angular momentum of the pair.
Therefore in this work we restrict ourselves to S, D, G pairs in $sd$ shell, and S, D pairs in $pf$ shell because of practicality.
This is also the starting point of the generalized-seniority scheme\cite{TALMI19711}, the interacting boson model \cite{Arima19751069}, and the Fermion Dynamic Symmetric Model \cite{WU1986313}.

We construct many-body basis states as
\begin{equation}
|\hat{A}^\dagger_{J_N M_N} \rangle = | ((\hat{A}^\dagger_{r_1} \otimes \hat{A}^\dagger_{r_2})_{J_2} \otimes \cdots \hat{A}^\dagger_{r_N} )_{J_N M_N} \rangle,
\label{NPA-bas}
\end{equation}
and calculate the Hamiltonian matrix elements
\begin{equation}
\langle \hat{A}_{J_N M_N} | \hat{H} | \hat{A}^\dagger_{J^\prime_N M^\prime_N} \rangle,
\label{NPA-hammtx}
\end{equation}
where $\hat{H}$ is as defined in (\ref{ham}), i.e., we use the general shell model 1+2-body interactions.
The basis kets defined in (\ref{NPA-bas}) are not orthogonal. 
As a result, the matrix defined in (\ref{NPA-hammtx}) has to be linearly transformed before diagonalization, and the linear transformation can be constructed from the overlap matrix $\langle \hat{A}_{J_N M_N} | \hat{A}^\dagger_{J^\prime_N M^\prime_N} \rangle$.

\begin{table}
\centering
\begin{tabular}{ccc}
\hline \hline
Nuclide & $D_{\rm NPA}$ & $D_{\rm FCI}$ \\ 
\hline
$^{52}$Fe~~~ & ~~~350~~~ & ~~~$1.1 \times 10^8$ \\ 
$^{53}$Fe~~~ & ~~~6106~~~ & ~~~$2.2 \times 10^8$ \\ 
$^{54}$Fe~~~ & ~~~706~~~ &  ~~~$3.5 \times 10^8$ \\ 
$^{56}$Fe~~~ & ~~~1276~~~ &  ~~~$5.0 \times 10^8$\\ 
\hline \hline
\end{tabular} 
\caption{ Typical $M$-scheme dimensions of Fe isotopes in NPA, compared with FCI (full configuration-interaction).}
\label{tab:NPADim}
\end{table}

This methodology is described in detail in \cite{ZHAO20141}. 
There are different versions of the nucleon-pair approximation \cite{PhysRevC.88.061303,PhysRevC.83.034321, Ginocchio19951861}, although this list is not exhaustive. 

While previous  implementations of the nucleon pair approximation utilized the natural $J$-scheme formalism \cite{CHEN1997686,ZHAO20141},
recent efforts have recast the NPA efficiently into the $M$-scheme \cite{PhysRevC.102.024304}.  Although the dimensionality of the $M$-scheme Hamilonian is 1-2 
orders of magnitude larger than in the $J$-scheme, the matrix elements themselves take considerably more time to compute in the latter due to a proliferation of 
recursion branches arising from angular momentum recoupling. Therefore, as in our FCI code \cite{BIGSTICK,johnson2018bigstick}, we adopt as more practical an $M$-scheme 
NPA approach, similar to that presented in \cite{PhysRevC.102.024304} but further simplified. 
Because the many-body Hamiltonian matrix elements in (\ref{NPA-hammtx}) are constructed by contraction of pair annihilation and creation operators via commutation relations \cite{CHEN1997686}, leading to  multiple recursions  in the code, the computation time increases exponentially with the number of valence particles. 
Our current limit is 10 valence protons and 10 valence neutrons which, with a 
 general $1+ 2$-body shell model interaction, takes less than 5 hours on a local 24-core workstation.

The virtue of the NPA is the small dimension of the truncated space. 
We present typical dimensions of several Fe isotopes in Table \ref{tab:NPADim}, in comparison with dimensions of FCI.
With such dimensions, the Hamiltonian can be easily diagonalized with standard Householder algorithm, and larger dimensions can be solved with  the Lanczos algorithm.
The time complexity of one basis overlap in NPA is proportional to $O(\Omega^2)$, where $\Omega$ is total proton/neutron capacity of a major shell, i.e. 12 for the $sd$ shell, 20 for $pf$ shell,
and so on. 
Therefore the NPA may be useful in generating the low-lying spectra of medium-heavy or heavy nuclei, for example in the $2s$-$1d$-$0g_{7/2}$-$0h_{11/2}$ space for which 
$\Omega = 32$.

\section{Results}

\begin{figure}\center
\includegraphics[width = 0.9 \textwidth]{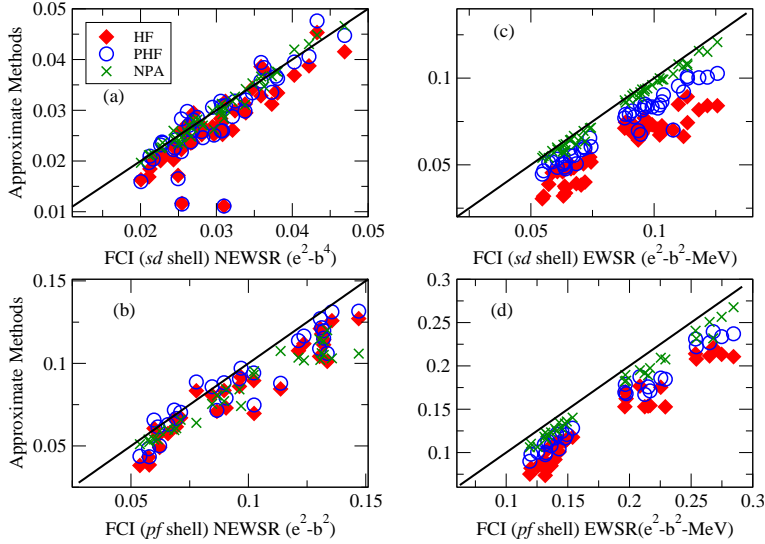}
\caption{\label{E2} Scatter plot of sum rules for  E2 transitions in the $sd$ shell, comparing the FCI results, which are numerically exact, 
against Hartree-Fock (red diamonds), projected Hartree-Fock (blue circles), and nucleon pair approximation NPS (green $x$s). Left-hand side, panels (a) and (b), are non-energy-weighted sum rules,
while right-hand side, panels (c) and (d), are energy-weighted sum rule. Upper panels (a) and (c) are from the $sd$ shell, while the lower panels (b) and 
(d) are from the $pf$ shell.  The diagonal straight line denotes perfect agreement.  See text for discussion.
}
\end{figure}

We considered three transitions: electric quadrupole (E2), magnetic dipole (M1), and Gamow-Teller. Because we were comparing 
 approximate methods against FCI calculations, all using the same operators (and not comparing to experiment), our results are not sensitive to scaling factors.
In the $sd$ shell, we computed all nuclides with $ 18 \geq N \geq Z \geq 10$. In the $pf$ shell we computed $^{44-52}$Ti, $^{46-54}$V, $^{48-53}$Cr, 
$^{50-55}$Mn, and $^{52-54,59-61}$Fe. Due to the challenges of odd numbers of particles, our NPA calculations did not include $^{52,54}$V, $^{53}$Cr,
or $^{52,54}$Mn. 

The E2 transition operator is \cite{BG77}
\begin{equation}
\hat{\cal O}(E2) = \sum_i e(i) r_i^2 Y_{2m}(\theta_i, \phi_i) \label{E2op}
\end{equation}
where the sum is over valence nucleons, with $e(i)$ the effective charge of the $i$th nucleon, and 
$r_i, \theta_i, \phi_i$ the spherical coordinates of the $i$th particle; $Y_{\ell m}$ is a spherical harmonic.  We  used effective charges of +1.5$e$ and +0.5$e$ for protons and 
neutrons, respectively, and assumed harmonic oscillators wave functions. By using the approximate formula for the harmonic oscillator frequency, $\hbar \omega 
\approx 41 A^{-1/3}$ MeV, where $A$ is the mass number, we chose for the $sd$ shell calculations $\hbar \omega = 13.5$ MeV (corresponding to $A=28$, and 
also leading to an oscillator length parameter $b = \sqrt{\hbar/m\omega} = 1.74$ fm), and 
for the $pf$ shell  we chose $\hbar \omega = 11.1$ MeV and $b= 1.93$ fm, corresponding to $A \sim 50$.  Again, we emphasize that because we are only benchmarking approximations, the overall scale is not deeply meaningful here. 
We compare our results in Fig ~\ref{E2} in scatter plots, comparing the FCI shell model calculations ($x$-axis) against our three approximations ($y$-axis), separating 
out the NEWSR, Fig.~\ref{E2}(a) and \ref{E2}(b) from the EWSR, Fig.~\ref{E2}(c) and \ref{E2}(d), as well as those from the $sd$ shell, Fig.~\ref{E2}(a) and \ref{E2}(c), 
and from the $pf$ shell, Fig.~\ref{E2}(b) and \ref{E2}(d). 
Perfect agreement would fall along the diagonal line.

\begin{figure}\center
\includegraphics[width = 0.9 \textwidth]{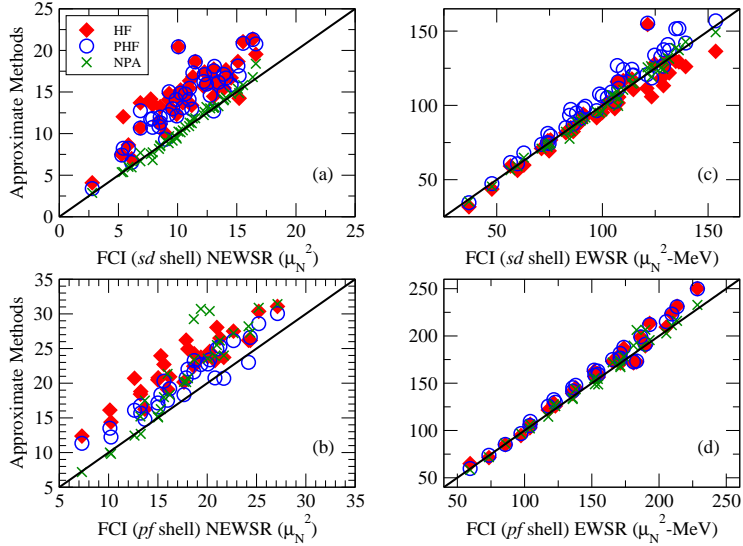}
\caption{\label{M1} Same as Fig.~\ref{E2}, but for M1 transitions}
\end{figure}

Two outlying sets of points for the  $sd$-shell E2 NEWSR are evident in Fig.~\ref{E2}: these are the HF and PHF approximations to $^{32,33}$S, which 
severely underestimate the NEWSR relative to FCI.  The HF state for $^{32}$S has 
filled $0d_{5/2}$-$1s_{1/2}$ shells and thus the PHF has only a single $J=0$ state, while the $^{33}$S HF minimum is almost entirely ($96\%$)  a single $0d_{3/2}$ neutron outside the filled 
$0d_{5/2}$-$1s_{1/2}$ shells.  Such spherical cases underestimate the deformation and thus the sum rules.  What is surprising is not so much the existence of these cases but rather that we found few other 
similar cases; for example, the HF minimum for $^{28}$Si is not spherical but slightly oblate. 
 We also see that, unsurprisingly, the NPA results generally give the best agreement, {if somewhat low for $pf$-shell cases}, followed by PHF, with HF yielding the relatively worst agreement.  On the other hand, 
even for HF the agreement is nonetheless relatively good, if systematically low.

The clumping evident in the E2 EWSR is by $Z$: in the $sd$ shell, Fig.~\ref{E2}(c), the lower-left clump is comprised of Ne, Na, Cl, and Ar isotopes, while the 
rest, Mg, Al, Si, P, and S, are in the upper-right clump; 
in the $pf$ shell, Fig.~\ref{E2}(d), the three clumps are for Ti and V,  Cr and Mn, and Fe, respectively.  The clumping arises because the effective charge for protons is, of course, larger than that 
for neutrons, and so the sum rules are dominated by expectation value of the purely proton part of the operators, and, as is well-known, the largest deformations are 
found in midshell.   Although in the E2 NEWSR plots there is no clumping evident by eye, the larger values also accrue to mid-shell values of $Z$.  We did not sample throughout 
the $pf$ shell because of computational cost, and because in the upper $pf$ shell the $0g_{9/2}$ should be included.
{All three approximations are slightly low for the $pf$ shell cases.}

\begin{figure}\center
\includegraphics[width = 0.9 \textwidth]{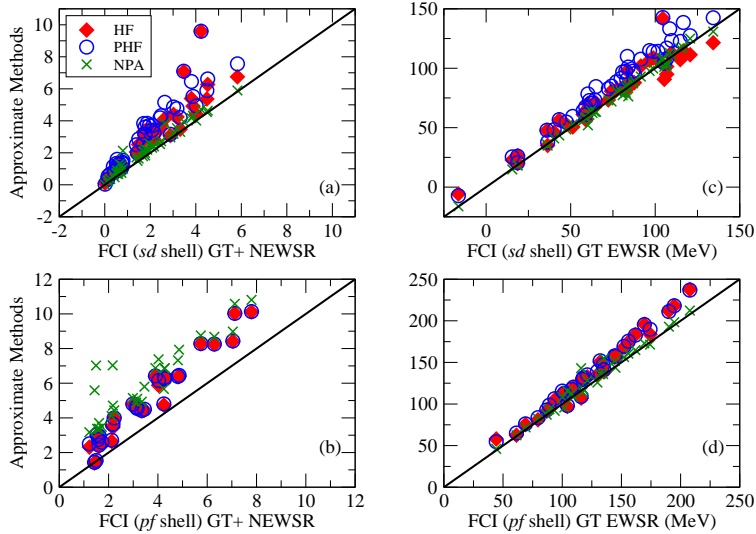}
\caption{\label{GT} Same as Fig.~\ref{E2}, but for Gamow-Teller transitions. The NEWSR is for $GT+$ only, while the EWSR includes both $GT+$ and $GT-$.}
\end{figure}

The magnetic dipole transition operator operator is \cite{BG77}
\begin{equation}
\hat{\cal{O}}(M1) 
=   \sqrt{\frac{3}{4\pi}} \sum_i \left ( g_s(i) \vec{s}_i + g_l(i) \vec{l}_i \right ),
\end{equation}
where we used  `bare' values for the $g$-factors: $g_l = 1, 0$ for protons and neutrons, respectively, while $g_s =  5.586, -3.826$ for protons and neutrons, 
respectively, all 
in units of the {nucleon magneton}, $\mu_N \equiv {e \hbar}/{2M_p c}$.
The M1 results are given in Fig.~\ref{M1}.
 Again for the $sd$-shell, $^{32}$S HF and PHF values are outliers, overestimating the M1 strength. Here the defect is not deformation but spin (although these are 
 related \cite{FRAZIER19977,PhysRevC.96.044319}): the expectation value 
 of total $\vec{S}^2$ is larger when filling the $0d_{5/2}$ and leaving the $0d_{3/2}$ empty, than for the FCI solution, which partially restores spin symmetry.

Additionally, in the $pf$ shell three NPA cases, $^{59,60,61}$Fe, provide notable
 overestimates of the M1 NEWSR.  Because the relevant valence neutron numbers are $13, 14, 15$ respectively, we perform a particle-hole or Pandya transformation \cite{BG77,lawson1980theory}, recasting  these nuclei into a proton-particle, neutron-hole representation, resulting with $7,6,5$ neutron holes respectively. Such transformations are common in configuration-interaction methods.  In our NPA calculations the neutron-hole pairs likely need further optimization, a topic for future work.

Finally, the  Gamow-Teller operator is $g_A \vec{\sigma} \tau_\pm$, where $g_A$ is the axial vector coupling constant, $\sigma$ is the Pauli matrix for spin, 
and $\tau_\pm$ is an isospin raising/lowering operator. Because for nuclei in these regions $g_A$ is quenched to a value of nearly 1, we simply took $g_A = 1$.
For the NEWSR we only consider $\beta^+$ transitions, as $\beta^-$ is related by the Ikeda sum rule, NEWSR(GT-) -NEWSR(GT+)$=3(N-Z)$, satisfied by all of our methods.
Our Gamow-Teller EWSR is actually the sum of the $\beta^+$ and $\beta^-$ transitions, because computing the EWSR operator requires it.  
In most of the nuclei we considered, ($Z \leq N$), the GT EWSR is dominated by $\beta^-$ which has a total strength $\approx 3(N-Z)$, so that  the  overall excellent agreement of GT EWSR is attributable to $\beta^-$ strengths.
Results are shown in Fig~\ref{GT}.   We have the same pattern of overestimation as for M1: $^{32}$S HF and PHF in the $sd$-shell, and $^{59,60,61}$Fe NPA for the 
$pf$ shell.  Prior work showed that smearing of the Fermi surface through deformation decreases the total Gamow-Teller strength, while conversely spherical states 
overestimate the Gamow-Teller strength \cite{PhysRevC.69.024311}. This is related to the overestimation of the M1 strengths.

\begin{figure}\center
\includegraphics[width = 0.7 \textwidth, clip]{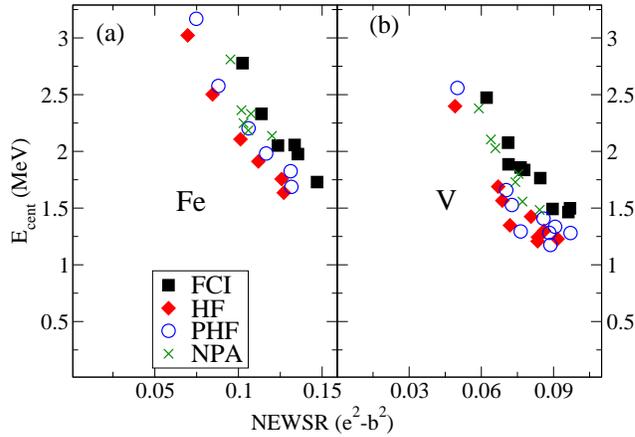}
\caption{\label{corre2} { Correlations between the E2  NEWSR and the centroid $E_\mathrm{cent}$, which is the ratio of the EWSR to the NEWSR, for (a) iron  and (b) vanadium isotopes.}}
\end{figure}

{
\subsection{Sample application: correlations in E2 sum rules}

\label{correlations}

Here we present a sample application.   As deformation increases, one expects the E2 transition strengths to increase, and the relative energy of quadrupole excitations, 
seen as $2^+$ levels in even-even nuclei, to decrease; such systematics are well known experimentally and well-understood \cite{casten2000nuclear}.  We compute the centroid of transition strength function, which is  the ratio of the EWSR to the NEWSR, and is simply the average excitation energy weighted by B(E2). 
\red{Although one should not assume that the 
centroid matches the \textit{peak} of a giant resonance, especially in the case of E2 resonances \cite{lipparini1989sum},}
such calculation can be used nonetheless to characterize  \red{ responses, for example the electric dipole (E1)} resonance \cite{PhysRevC.59.3116} (those authors define their moments with integrals over cross-sections, not strength functions, 
so their $\sigma_\mathrm{int}$ is proportional to our EWSR and their $\sigma_{-1}$ is proportional to our NEWSR).

  In Fig.~\ref{corre2} we show the expected anti-correlation between the NEWSR, or total strength, versus the centroid energy, for iron (left panel, \ref{corre2}(a)), and vanadium (right panel, \ref{corre2}(b)) isotopes.   \red{We point out these energies are more typical of low-lying E2 transitions and not giant resonances, which cannot be  described in a fully 
  consistent manner
 in a single harmonic oscillator shell, as the E2 operator (\ref{E2op}) connects to other shells}.
    Our point here is that our approximate methods all show the same trends and even the same slope, if with different offsets, from 
the ``exact'' FCI calculation. Furthermore, we can carry out such a study for odd-$A$ and odd-odd nuclides and find the same trend replicated.   With these tools it is easy to explore for
unexpected correlations, using simpler approximations, inspiring more detailed  investigations.

}

\section{{Analysis}}

\label{analysis} 

We can interpret our results in light of previous investigations on the Brink-Axel hypothesis. The Brink-Axel hypothesis postulates that strength functions off excited states have the same dependence 
on $E_x = E_f - E_i$ as the ground state, which would imply that sum rules are also the same.  
Indeed, as a partial test of the Brink-Axel hypothesis, numerical experiments have investigated both the NEWSR \cite{johnson2015systematics} and EWSR \cite{PhysRevC.97.034330}, 
using the same FCI framework as above.  These investigations showed that the sum rules, while not constant, exhibit a steady secular evolution with $E_i$, which can be understood 
mathematically and agree within a robust 
statistical fluctuation \cite{johnson2015systematics}.  This means that states that are nearby in energy will have similar sum rules; {furthermore, if one puts the sum rules into energy 
bins and computes the average and standard deviation in those bins, the standard deviation (which represents the fluctuations) is not sensitive to the bin size}.

\begin{figure}\center
\includegraphics[width = 0.9 \textwidth, clip]{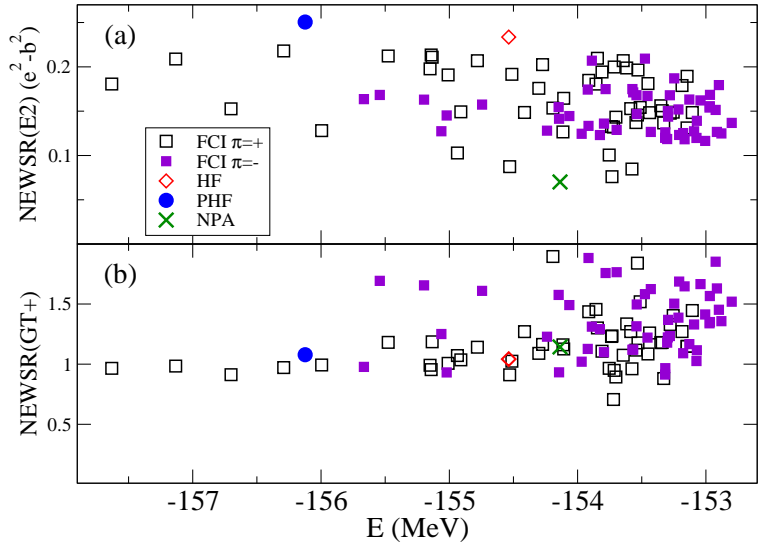}
\caption{\label{sr} {NEWSR for E2, panel (a), and GT+, panel (b), for $^{76}$Sr, for low-lying positive and negative parity FCI states, and for (positive-parity) 
HF,  PHF, and NPA states.
\red{The energy $E$ is that of the individual FCI, HF, PHF, or NPA states.}
 }}
\end{figure} 

{We explored this in detail by  
computing the E2 and the GT+ NEWSR for $^{76}$Sr in the space of $1p_{1/2}$-$0f_{5/2}$-$0g_{9/2}$ using the JUN45 interaction 
\cite{PhysRevC.80.064323} 
(to aid in calculation, we truncated the space by filling the $1p_{3/2}$ orbit). 
For E2 we chose the same effective charges as previously, and used an oscillator frequency of 9.68 MeV and thus an oscillator length of 2.07 fm; for GT, as previously, we used 
an effective $g_A \approx 1$. 
  In this space we  compute in FCI the NEWSR for the lowest fifty states,  for both positive and negative parities, as well as the HF, PHF, and NPA (computed only with 
  SD pairs) values; our HF state did not mix parity. 
  The NPA energy is higher than our HF energy, but could be improved by including pairs of higher angular momentum.  Fig.~\ref{sr}, 
  \red{plotting sum rules against the energies of individual states}, 
  demonstrate a number of points.  First, one can see 
 the NEWSR are similar for individual \red{low-lying FCI states},  albeit with fluctuations, as discussed elsewhere \cite{johnson2015systematics}.  The systematics are slightly different from positive and negative 
parity states. 

Second, while the HF and PHF NEWSR is well within the range of FCI values for GT+, for our HF and PHF E2 they are actually above the local values, while 
the NPA E2 value is below. This can be understood by expanding 
any approximate state, $| \Psi_A \rangle$, in the eigenstates $| i \rangle$ of the Hamiltonian:
\begin{equation}
| \Psi_A \rangle = \sum_i c_i | i \rangle.
\end{equation}
Then, for a sum rule operator $\hat{\cal O}_\mathrm{SR} $,  the evaluated sum rule is 
\begin{eqnarray}
\langle \Psi_A | \hat{\cal O}_\mathrm{SR} | \Psi_A \rangle = 
\sum_{i,j} c_i^* c_j \langle i | \hat{\cal O}_\mathrm{SR} | j \rangle \nonumber \\ 
= \sum_i |c_i|^2 \langle i | \hat{\cal O}_\mathrm{SR} | i \rangle + \sum_{i \neq j} c_i^* c_j \langle i | \hat{\cal O}_\mathrm{SR} | j \rangle 
\label{sumruleexpansion}
\end{eqnarray}
One can see the sum rule consists of a coherent sum, which is just a weighted average over the sum rules of the components of the approximate state, 
and an incoherent sum over off-diagonal matrix elements. If we had only the coherent sum, we would be guaranteed that the sum rule on an 
approximate state is simply a weighted average of its components.  The off-diagonal contributions, however, spoil this nice picture, as demonstrated 
in the E2 results in Fig.~\ref{sr}.  On the other hand, one might expect for random signs in the off-diagonal contributions to tend to cancel.  Therefore, although
we lack a theorem, we have an explanation why our results track the actual ground state sum rules. 
The sum rules are largely an average of sum rules from low-lying states, the diagonal terms in Eq.(\ref{sumruleexpansion}), which in turn have similar sum rules,

}

This is good news, because it means exactly what one hopes for: that even approximate ground states are {reasonable} proxies for the ground state sum rules.

\section{Conclusions}

We have compared non-energy-weighted and energy weighted sum rules for several important transition operators, evaluated by taking expectations values in both numerically exact and 
several approximate ground state wave functions.   We found general good agreement, with improved agreement  naturally linked to more sophisticated approximations.  The good agreement
can be understood within the context of the Brink-Axel hypothesis: while eignenstates distant in energy can and do have very different sum rules, eigenstates nearby in energy tend to have 
similar sum rules. The approximate ground states are superpositions of true eigenstates, but dominated by states near the true ground state, and hence sharing similar sum rules. 
Our final conclusion is that approximate ground states can be useful in learning something about exact sum rules, and thus about nuclear dynamics.

\section{Acknowledgements}

This material is based upon work supported by the U.S. Department of Energy, Office of Science, Office of Nuclear Physics, 
under Award Numbers  DE-FG02-03ER41272.  
CWJ and YL are grateful to the CUSTIPEN (China-U.S. Theory Institute for Physics with Exotic Nuclei) program, funded by the U.S. Department of Energy, Office of Science under grant number DE-SC0009971, which initiated this collaboration.
YL acknowledges support by National Natural Science Foundation of China (Grant No. 11705100), the Youth Innovations and Talents Project of Shandong Provincial Colleges and Universities (Grant No. 201909118), Higher Educational Youth Innovation Science and Technology Program Shandong Province (Grant No. 2020KJJ004).

\appendix

\section{Computational resources}

The {\tt BIGSTICK} shell model configuration interaction code is available as free open-sources software \cite{BigstickCode}. 
At that same site are tools for generating the one-body matrix elements for the transition operators used.  The {\tt PandasCommute} code \cite{PandasCommute} was used to 
generate the sum rule matrix elements.

At this time, the Hartree-Fock and angular momentum projection codes used ({\tt SHERPA} and {\tt LAMP}, respectively) are not publically available, nor the NPA code.

\bibliographystyle{unsrt}

\bibliography{johnsonmaster,NucleonPairApproximation}

\end{document}